# Quantum Control with the Level Set Method

**Fariel Shafee**
Physics Department
Princeton University
Princeton, NJ 08540

*Abstract:*

We examine the relevance of the Level Set Method (LSM) in coherent control of quantum systems where the objective is to retain or attain a particular expectation value of a given measurable. The differences with the usual applications of LSM, where continuous closed interfaces are involved, and the quantum case, where we may have a discrete number of points to deal with, are noted. The question of optimization in this new context is also clarified. Simple examples with symmetric and asymmetric multidimensional potentials are briefly considered.

## 1. Introduction

In recent years the Level Set Method (LSM) [1-3] has been used in a wide range of classical situations where an interface between two different phases of matter is seen to evolve in time due to dynamic factors, which may also involve the complications of curvature and entropy. The speed of motion of the interface contour is usually nonuniform in the general case and one has to deal with the motion of a finite, in fact reasonably small, set of marker points on the contour in place of the infinite number of points that define the curve. This makes the representation of the evolution a highly nontrivial problem. Geometrical changes in the contours with regimes of high curvature, acnodes and cusps may be smoothed with a number of techniques developed and even topological changes, once highlighted by Rene Thom's catastrophe theory [4], may arise in certain situations, simply from a study of the right model for numerical time evolution. This makes this a fascinating technique and points out the possibility of its relevance also in non-classical contexts where moving fronts are involved.

We have previously studied [5] the quantum control of the expectation value of a measurable in a quantum system. If the measurable is the energy itself or a function of the energy, then one can in principle find a sequence of unitary operators, representing a sequence of laser pulses, that can take an arbitrary initial state to a desired state with the desired expectation value as described in detail by Ramakrishna, Rabitz, Schirmer et al. [6]. In case the variable measured does not commute with the Hamiltonian and its eigenstates form a different basis, we have shown how, in certain simple cases where variations in the expectation value of the designated measurable due to transitions involving only a small number of energy eigenstates, the problem can again be tackled by a series of pulses to restore the original state correcting the deviation in a coherent way.

In this work we shall investigate the problem from the perspective of the LSM. However, we must first be aware of the differences of the classical and the quantal contexts.

## 2. Level Set Method in Fluid Motion

If a chosen interface in a material is represented by a closed contour described by:

$$\phi(p,t) = 0 \qquad (1)$$

where $p$ represents the coordinates of a material particle at time $t$, and $\phi$ is simply the signed distance of the particle at time $t$ from its position at time $t = 0$, then we get the convective equation:

$$\partial \phi / \partial t + \nabla \phi . u(p) = 0 . \tag{2}$$

where $u$ is the velocity of the particle $p$.

The velocity $u$ is modeled according to the context. Obviously $u.\nabla \phi$ is the speed of particle normal to the contour, it is signed with opposite for incoming or outgoing (converging or diverging) motion. It may have a constant part and other components depending on entropy and curvature.

Since one always works with a finite grid of points, the method involves an algorithm for identifying a level set of grid points which best approximates the contour. $\phi$ cannot be zero at a grid point except as an accident. But working one's way in starting at the outside boundary one can split the grid points into two parts – those completely outside the contour which have no neighbors inside the contour, i.e. for which the distance to all points [approximated by a finite number] of the curve is greater than the distance to its neighbors, and those for which this test fails. The failed points of the grid form the interface.

Equation 2 can then be used with an appropriate formulation of the speed function for progress of the interface contour.

When two straight portions of the curve meet at an angle, the common point creates a singularity as it gets two different velocities. Hence it is necessary to introduce shock waves, i.e. entropy considerations to smooth out the kink.

A similar LSM has been used also for reconstruction of three-dimensional images from sets of two-dimensional data. Here $t$ becomes the third dimension $z$. So a set of points in three dimensions is interpolated to form a continuous surface of various degrees of coarseness according to a scheme best suited for the specific type of geometry.

## 3. Quantum Evolution of Wave Functions and LSM

In optics the dynamics of wave fronts can often be generated by the eikonal approximation with local gradients of the fronts giving the direction of forward development. Cheng [7] has recently pointed out the possibility of having similar evolution of the quantum wave function as dictated by Schrodinger equation. It may be noted here that the wave function is a complex quantity, and hence the real and imaginary parts must behave differently. Indeed by the Cauchy-Riemann equations they form orthogonal contours. One may use modulus contours and phase contours of the amplitude for a coupled evolution. The modulus contours can be expected to be closed curves around the singularities and the zeros of the wave function, and the phase contours would then be orthogonal families of curves connecting singularities and zeros.

## 4. Expectation Values in State Space

In all the previous instances of the use of the LSM, the method dealt with real co-ordinate space. However, we come across surfaces also in parameter space and hence LSM may be a relevant technique to approximate functional evolution also in parameter space in certain cases. Let us consider the case of the expectation value of energy, which is the simplest. Let the system have only three eigenstates with energies $E_1$, $E_2$ and $E_3$. Let an arbitrary state be expanded

$$|\psi> = \sum_i a_i |E_i> \qquad (3)$$

with

$$<E> = \sum_i |a_i|^2 E_i \qquad (4)$$

But on account of the normalization condition we have only two independent $a_i$:

$$\sum_i |a_i|^2 = 1 \sum_i |a_i|^2 = 1 \qquad (5)$$

If we choose for simplicity real coefficients (which is possible for bound states), then we get the elliptical contours in $(a_1, a_2)$ plane:

$$a_1^2 (E_1 - E_3) + a_2^2 (E_2 - E_3) = <E> - E_3 \qquad (6)$$

assuming $E_3$ is the minimum eigenvalue.

Hence different $<E>$ values create different level sets. If we want the system to proceed to a particular $<E>$ from any other $<E>$, we have to use unitary transformations to perform the job. A state is again given by a particular point $|p> = (a_1, a_2)$ in this coefficient space, and its trajectory from one contour to another is similar to fluid convection, with unitary operators generated possibly by laser pulses acting as the driving force.

However, an interesting dissimilarity is that if we are interested in only attaining a particular $<E>$, then any unitary operator that takes $|p>$ to the target contour may suffice. So it is unnecessary to have normal linear flow or to apply viscous damping forces or entropy increasing terms to smooth out the trajectories.

However, we may also be interested in taking an ensemble of systems with the same $<E>$ to a different $<E>$ in which case we may be interested in studying the evolution of the total level set and not just a single state. In this case we need a sequence of pulses for each grid point on the level set, and the totality of the set of these unitary transformations define the motion of the contour.

The distribution of matter in real space involves to some extent the question of incompressibility or at least limited compressibility and hence an ensemble of particles in real space cannot be compressed to a single point. On the other hand in the quantum context it is possible to make all points on a level set map onto a single point on another level set through different trajectories, some parts of which may be common, unless the exclusion principle prevents it in a fermionic ensemble. A simple way would be to bring all the states on the level set to a particular point on the same level set by a similar set of unitary operations

$$U(p_o, p)|p> = |p_o>  \quad (7)$$

where $|p_o>$ is the anchor target, and $|p>$ are all states of the $<E>$ contour. Obviously there is a diffeomorphic mapping between the parameters $(a_1, a_2)$ and the elements of the unitary operator $U$. We have previously proved that it is impossible to have a single $U$ taking all $|p>$ to $|p_o>$.

After all states are taken to $|p_o>$, it is a relatively simple task to design a unique sequence of laser pulses that takes $|p_o>$ to a single state (point) of the final target contour, and hence from the second anchor it can be redistributed by another set of unitary transformations to all members of the second level set.

$$U(q_o, p_o)|p_o> = |q_o>  \quad (8)$$

$$U(q, q_o)|q_o> = |q>  \quad (9)$$

What we have indicated symbolically as the relevant unitary operators in the equations above can be made more concrete in the context of a specific Hamiltonian by methods detailed in the work of Ramakrishna, Rabitz, Schirmer et al.

This three-step method of contraction of a level set to a point on the level set, transfer to another level set and then re-expansion to points in the second level set, is a uniquely quantum possibility not likely to be useful in any classical situation. However, in terms of designing the required unitary operators, and hence the experimental sequence of pulses, either square, or Gaussian, they may be more convenient. Hence the question of optimization would involve the definition of the cost function, and would be context-sensitive.

On the other hand we can also use Eq. (6) to make direct point to point transitions through bijective unitary transformations. One can use the classical LSM to locate a unique target point on the second level set for each initial point on the first level set and then find the unitary map to make the transition:

$$|q> = U(q, p)|p>  \quad (10)$$

One can rewrite Eq. (6) in the matrix form

$$a^T E a = 1 \tag{11}$$

where

$$E = \begin{bmatrix} E'_1 & 0 \\ 0 & E'_2 \end{bmatrix} \tag{12}$$

with

$$E'_{1,2} = (E_{1,2} - E_3)/(<E> - E_3) \tag{13}$$

Hence for a small change

$$da^T E a + a^T E\, da = a^T dE\, a \tag{14}$$

where the change in the matrix $E$ comes from the change in the $<E>$.

Eq. 14 gives the "equation of motion" of the level set.

When the measured quantity $\theta$ is not the energy, the problem is somewhat more complicated. This would require a change in basis as we have discussed in earlier work. On the $(a_1, a_2)$ energy state space a constant $<\theta>$ contour will in general form different contours which may not even be closed. It is interesting to note that since two curves intersect at a finite number of points the energy level set will have a only a few points, possibly only one, or two, with a given $<\theta>$ and a given $<E>$. In this case we may simplify the problem in finding the few U matrices taking the discrete $|p_i>$ to the discrete $|p_j>$ if $<E>$ is kept fixed, but $<\theta>$ is changed, or to discrete $|q_j>$ if both $<E>$ and $<\theta>$ are changed.

## 5. Level Sets in Hamiltonian Parameter Space

In certain problems Hamiltonian parameters are related to real space. As in the case of state space, we can also have a metric associated with coordinates in this Hamiltonian parameter space and hence a concept of distance and the relevance of the LSM.

We consider the simplest case of a two-dimensional anisotropic oscillator:

$$E = (n_x + 1/2)\, \omega_x + (n_y + 1/2)\, \omega_y \tag{15}$$

where $x$ is a 2-dimensional vector and $K$ is a *2X2* symmetric matrix. We shall assume it to be diagonal, because even if it is not so originally it can be converted into one.

The equal $<H_{int}>$ contours are of course elliptical. Actually quantization can be performed for the two orthogonal directions separately, leading to the discrete spectra (Fig. 2):

$$E = (n_x + 1/2)\, \omega_x + (n_y + 1/2)\, \omega_y \qquad (16)$$

So equal $E$ sets are actually discrete points lying on straight lines in the $(n_x, n_y)$ grid, if $\omega_x$ and $\omega_y$ have a rational ratio. If the ratio is not rational, there is just one $|p>$ for each given $E$, i.e. there is no degeneracy. In the case of anharmonic oscillators or the Morse potential one can have unequal unique spacing of the energy levels even of a single dimension. This makes the problem of tuning the laser pulses uniquely for transitions between only two given states. However, the relevance of LSM becomes unclear in this case, as there the equal $E$ lines are populated effectively by just one point in the $(n_x, n_y)$ grid.

For the more interesting case of equal $\omega_x$ and $\omega_y$ (Fig. 2), the equal $E$ level sets, though still discrete on the $(n_x, n_y)$ grid points are close packed in a sense. As the value of $\omega_x = \omega_y$ is changed, the $E$ values are changed for the same $(n_x, n_y)$ point sets. So, the $E$ level sets move on the $(n_x, n_y)$ grid in a discrete fashion as $\omega_x$ changes by rational fractions or multiples. However, it is not obvious how a change in the parameters of the Hamiltonian of the system may be obtained by quantum control with an interacting Hamiltonian.

## 6. Conclusions

We have here concentrated on the clarification of general ideas related to the possible application of the LSM in quantum control, especially for expected energy of a system.

If instead of $E$ we are interested in a different variable $<\theta>$, then again we have to contend with a different set of contours in the $(n_x, n_y)$ grid, and also have to deal with superposition of states. This problem is now being studied as well as a detailed numerical study of two-dimensional Morse potentials with a finite number of unequal energy spacings.

The author thanks Professor H. Rabitz for useful discussions.

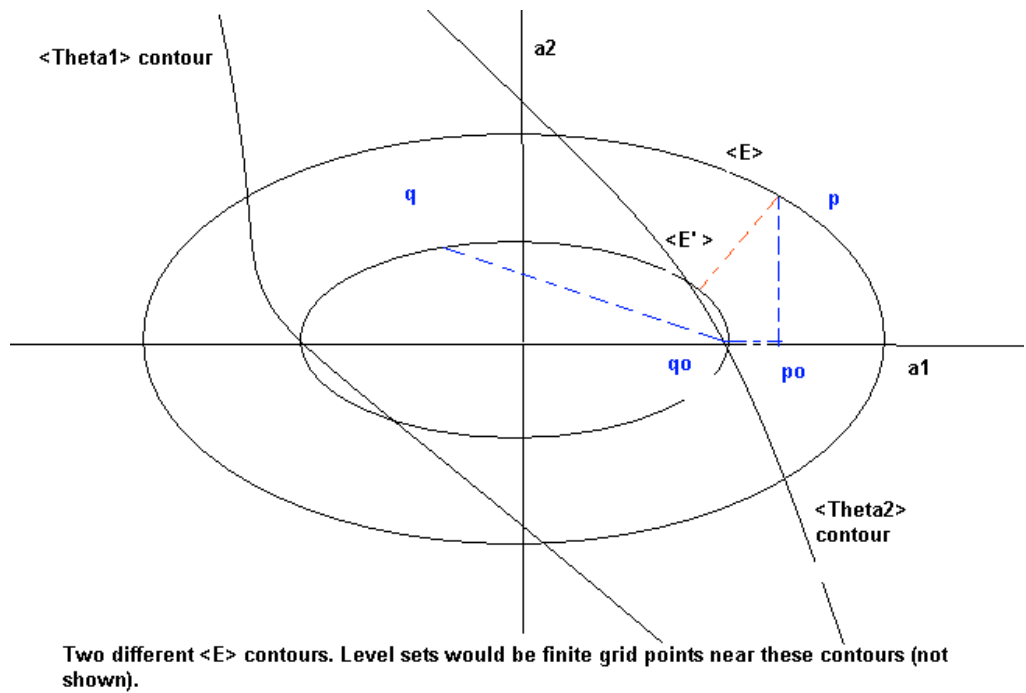

Fig. 1: Ellipses for different <E> in state space. Blue trajectory is the three-step transition discussed in the text, and the red trajectory is a one-step transition by rescaling. Hypothetical < θ> level sets differing from the <E> sets are also shown for comparison.

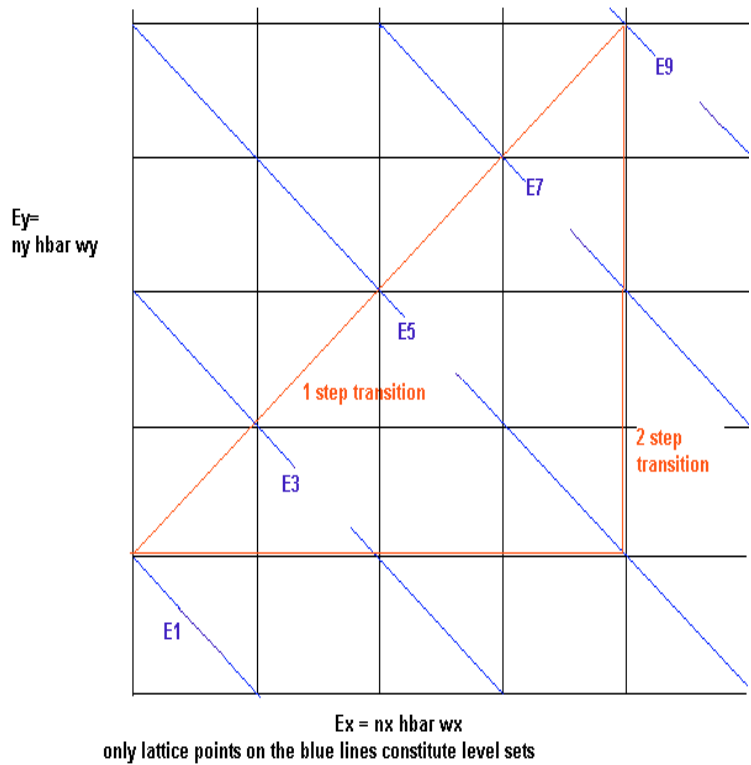

Fig 2: Level sets for a 2-dimensional harmonic oscillator, consists of discrete points on lattice meeting the corresponding set lines (we have omitted the zero-point energy for clarity). The transition from $E_9$ to $E_1$ may take place in many ways; the two simplest are shown by red lines. The two-step process involves laser pulses corresponding to a frequency $4\omega_y$ followed by $4\omega_x$.